\documentclass{article}
\usepackage{amsmath,graphicx,mlspconf}
\usepackage{cite}
\usepackage{amsmath,amsfonts}
\usepackage{algorithmic}
\usepackage{graphicx}

\usepackage{textcomp}
\usepackage{xcolor}

\usepackage{CJKutf8}
\usepackage{tabularray}
\usepackage{subfigure}
\usepackage{caption}
\usepackage{multirow}
\usepackage{bm}
\usepackage{amsmath,amsfonts}
\usepackage{algorithmic}
\usepackage{algorithm}
\usepackage{array}
\usepackage{xcolor}
\usepackage[caption=false,font=normalsize,labelfont=sf,textfont=sf]{subfig}
\usepackage{textcomp}
\usepackage{stfloats}
\usepackage{url}
\usepackage{verbatim}
\usepackage{graphicx}
\usepackage{cite}
\usepackage{amsthm}

\usepackage{amsmath,amsfonts}
\usepackage{algorithmic}
\usepackage{array}
\usepackage[caption=false,font=normalsize,labelfont=sf,textfont=sf]{subfig}
\usepackage{textcomp}
\usepackage{stfloats}
\usepackage{url}
\usepackage{verbatim}
\usepackage{graphicx}
\def\BibTeX{{\rm B\kern-.05em{\sc i\kern-.025em b}\kern-.08em
    T\kern-.1667em\lower.7ex\hbox{E}\kern-.125emX}}
\usepackage{balance}
\usepackage{amsmath}
\allowdisplaybreaks[4]
\begin{document}
\allowdisplaybreaks[4]
\newtheorem{myDef}{Definition}
\newtheorem{myPro}{Property}

\begin{CJK}{UTF8}{gbsn}
\toappear{2024 IEEE International Workshop on Machine Learning for Signal Processing, Sept.\ 22--25, 2024, London, UK}

\title{Enhancing Image Privacy in Semantic Communication over Wiretap Channels leveraging Differential Privacy}

\name{Weixuan Chen, Shunpu Tang, Qianqian Yang\thanks{Preprint submitted to IEEE MLSP 2024.}}
\address{College of Information Science and Electronic Engineering, Zhejiang University, Hangzhou, China}

\maketitle

\begin{abstract}
Semantic communication (SemCom) enhances transmission efficiency by sending only task-relevant information compared to traditional methods. However, transmitting semantic-rich data over insecure or public channels poses security and privacy risks. This paper addresses the privacy problem of transmitting images over wiretap channels and proposes a novel SemCom approach ensuring privacy through a differential privacy (DP)-based image protection and deprotection mechanism. The method utilizes the GAN inversion technique to extract disentangled semantic features and applies a DP mechanism to protect sensitive features within the extracted semantic information. To address the non-invertibility of DP, we introduce two neural networks to approximate the DP application and removal processes, offering a privacy protection level close to that by the original DP process. Simulation results validate the effectiveness of our method in preventing eavesdroppers from obtaining sensitive information while maintaining high-fidelity image reconstruction at the legitimate receiver.
%
%
%
%

%

\end{abstract}


\section{Introduction}

Semantic communication (SemCom) emerges as a promising technology to enhance communication efficiency, recognized as one of the key techniques in next-generation networks \cite{qin2021semantic}. SemCom focuses on transmitting the intended meaning or information of the data, thereby reducing the amount of irrelevant or redundant information transmitted between users. A pioneering work on SemCom system for image transmission, named \textit{deep joint source-channel coding} (DeepJSCC) was proposed in \cite{bourtsoulatze2019deep}, where the semantic information is extracted and reconstructed by deep neural networks (DNNs), demonstrating superiority over conventional digital communication system. Building upon this, numerous works have been proposed to further enhance the communication efficiency\cite{zhang2022wireless,wang2022wireless}.

While SemCom systems offer significant advantages, the open nature of wireless channels and the semantic-rich characteristics of transmitted signals have raised concerns regarding their security and privacy \cite{chen2024nearly}. 
To tackle this issue, the authors in\cite{erdemir2022privacy, marchioro2020adversarial} proposed adversarial training approaches to train semantic encoder and decoder by minimizing information leakage while ensuring image reconstruction performance at the receiver.
However, these works often assumed that the channel condition of the eavesdropping link is significantly worse than the link between the transmitter and the legitimate user. Additionally, they assumed that the legitimate user knows the decoder architecture of the eavesdropper, which may not be realistic in real-world scenarios. In \cite{tung2023deep}, the authors proposed encrypting the quantized semantic information through a public key cryptographic scheme without requiring any prior knowledge. However, this scheme is vulnerable to quantum computing attacks \cite{buchanan2017will}. Another approach in \cite{han2023generative} introduced a SemCom system by manipulating the private semantic information extracted by Semantic StyleGAN \cite{shi2022semanticstylegan} before transmission. Nevertheless, this method solely focuses on protecting image privacy and does not address the process of deprotecting and reconstructing the received image by the legitimate user.


In this paper, we propose a novel SemCom approach for privacy-preserving image transmission, offering controllable privacy guarantees. This approach extracts disentangled latent features from a source image using the GAN inversion method and then apply a differential privacy (DP) mechanism to specific latent features with privacy-preserving requirements. As the DP process is not invertible, we introduce two neural networks (NNs) to approximate the DP application and removal processes, thereby providing a privacy guarantee close to that of the original DP process. Simulation results demonstrate that the proposed approach effectively prevents privacy leakage to eavesdroppers by generating chaotic or fake images to mislead eavesdroppers, while maintaining high-fidelity image reconstruction for legitimate users compared to the scenarios without DP mechanisms.

\section{Preliminaries}


%
%

Differential privacy (DP) \cite{dwork2006differential} is a privacy-preserving mechanism inspired by indistinguishability in cryptography, given by

\begin{myDef}[$\epsilon$-Differential Privacy] 

Let $D$ and $D'$ be any neighboring datasets that differ on one element, and $y$ be any released output. 
A randomized mechanism $M$ gives $\epsilon$-differential privacy if it satisfies:
\begin{equation}
Pr[M(D) = y] \leq e^\epsilon \cdot Pr[M(D') = y].
\end{equation}
\end{myDef}
Here $\epsilon$ is the privacy budget, and the smaller the value of $\epsilon$, the stronger the privacy guarantee.
The method to achieve DP is to add a random perturbation to the data. 
Before perturbation, we need to determine the sensitivity, which is used to calibrate the amount of noise for a specified query $f$ of dataset $D$.
The $l$1-norm sensitivity in DP can be defined as:
\begin{myDef}[$l$1-norm Sensitivity \cite{dwork2006differential}]
Let $f: D \rightarrow \mathbb{R}$ be any query, $l$1-norm sensitivity $\Delta f$ satisfies the following equation:

\begin{equation}
\Delta f = \max_{D, D'} \left\| f(D) - f(D') \right\|_1.
\end{equation}
\end{myDef}


Random perturbation is usually generated based on the Laplace mechanism,
which can be defined as:
\begin{myDef}[Laplace Mechanism \cite{dwork2006differential}]
Given a function $f: D \rightarrow \mathbb{R}$, a mechanism $M$ gives $\epsilon$-differential privacy if it satisfies the following equation:

\begin{equation}
M(D) = f(D) + Lap\left(0, \frac{\Delta f}{\epsilon}\right).
\end{equation}
\end{myDef}

Here $Lap\left(0, \frac{\Delta f}{\epsilon}\right)$ represents a Laplace distribution with location parameter 0 and scale parameter $\frac{\Delta f}{\epsilon}$.



\begin{figure*}[htbp]
\begin{center}
\centerline{\includegraphics[width=0.8\linewidth]{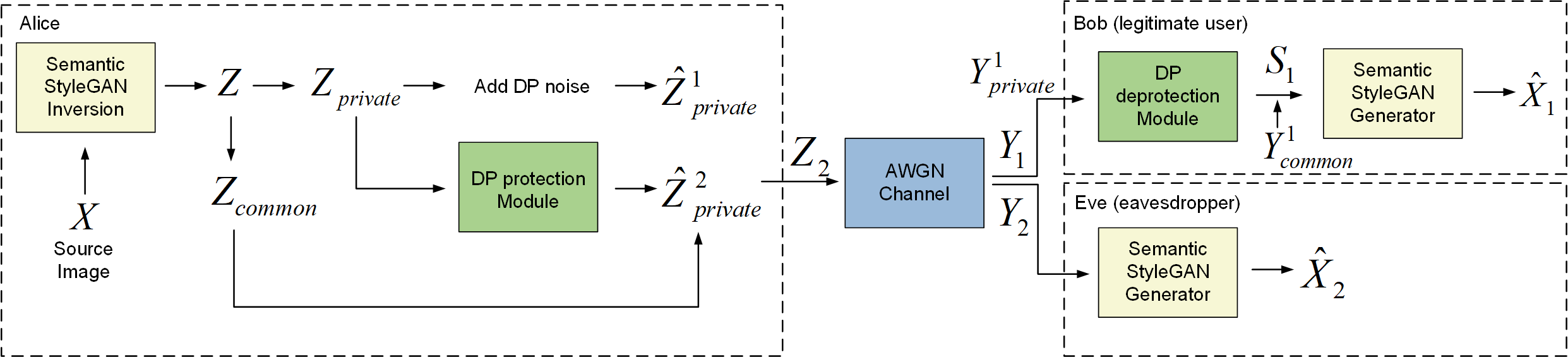}}
\caption{The framework of our proposed system.}
\label{fig.1}
\end{center}
\vskip -0.3in
\end{figure*}
\section{Problem Setup and System Design}

\subsection{Problem Setup}

As shown in Fig.~\ref{fig.1}, we consider a learning-based SemCom system under AWGN channels, where a transmitter aims to reliably transmit a private facial image to a receiver, while an eavesdropper attempts to recover this image.  We use $\textbf{X}$ to denote the source image. The legitimate sender, legitimate receiver, and eavesdropper are referred to as Alice, Bob, and Eve, respectively.
Alice utilizes the inversion method of a pre-trained semantic StyleGAN $f_{\mathrm{inv}}$ to transform $\textbf{X}$ into a disentangled latent representation $\textbf{Z}$. 
This latent representation contains several disentangled latent codes and can be regarded as important semantic information. Each latent code represents a specific attribute of the facial image, such as the eyes or nose.
Mathematically, this process can be written as
\begin{equation}
    \textbf{Z} = f_{\mathrm{inv}}\left( {\textbf{X}} \right).
\end{equation}
Then we apply DP mechanism to the private part of the latent codes. 
Specifically, we denote the private latent codes as $\textbf{Z}_{\rm private}$, and the remaining latent codes as $\textbf{Z}_{\rm common}$. We then have
\begin{equation}
    \textbf{Z} = \left[\textbf{Z}_{\rm private}, \textbf{Z}_{\rm common}\right].
\end{equation}
We add genuine DP noise to $\textbf{Z}_{\rm private}$ based on a given privacy budget $\epsilon$, denoted by
\begin{equation}
    \hat{\textbf{Z}}_{\rm private}^{1} = \textbf{Z}_{\rm private} 
    + \textbf{n}_{\rm dp},
\end{equation}
where $\textbf{n}_{\rm dp} \sim Lap(0, \frac{\Delta f}{\epsilon})$, and we will discuss how to obtain $\Delta f$ in the following.
Subsequently, we combine $\hat{\textbf{Z}}_{\rm private}^{1}$ and $\textbf{Z}_{\rm common}$ to obtain $\textbf{Z}_1$, denoted by
\begin{equation}
    \textbf{Z}_1 = \left[\hat{\textbf{Z}}_{\rm private}^{1}, \textbf{Z}_{\rm common}\right].
\end{equation}
However, directly adding DP noise to $\textbf{Z}_{\rm private}$ would make it difficult for the legitimate user to remove the noise. 
Therefore, we use a NN to add fake DP noise, allowing the legitimate user to employ another NN to remove this noise.
Specifically, $\textbf{Z}_{\rm private}$ is fed into a NN-based DP protection module to obtain the perturbed semantic information $\hat{\textbf{Z}}_{\rm private}^{2}$, denoted by
\begin{equation}
\hat{\textbf{Z}}_{\rm private}^{2} = f_{\mathrm{protection}}\left( {\textbf{Z}_{\rm private};\bm{\theta}^{\mathrm{protection}}} \right),
\end{equation}
where $f_{\mathrm{protection}}$ represents the NN-based DP protection module.
The objective of the NN-based DP protection module is to make $\hat{\textbf{Z}}_{\rm private}^{2}$ as similar as possible to $\hat{\textbf{Z}}_{\rm private}^{1}$.
We combine $\hat{\textbf{Z}}_{\rm private}^{2}$ and $\textbf{Z}_{\rm common}$ to obtain the semantic information $\textbf{Z}_2$, denoted by
\begin{equation}
    \textbf{Z}_2 = \left[\hat{\textbf{Z}}_{\rm private}^{2}, \textbf{Z}_{\rm common}\right].
\end{equation}
Note that the semantic information we actually transmit is $\textbf{Z}_2$, and the role of $\textbf{Z}_1$ is to guide the generation of $\textbf{Z}_2$ based on the given privacy budget $\epsilon$.

$\textbf{Z}_2$ is then power normalized by $P$ and converted to a complex vector, which is subsequently transmitted over the AWGN channel to Bob.
Bob receives the noisy semantic information 
\begin{equation}
  \textbf{Y}_{1} = \textbf{Z}_2 + \textbf{n}_1,   
\end{equation}
where $\textbf{n}_1 \sim \mathcal{CN} (0, {\sigma^{2}} )$.
In this paper, we assume that the eavesdropper has the same channel condition as the legitimate user. 
Eve eavesdrops on the transmitted semantic information through her own AWGN channel, denoted by
\begin{equation}
  \textbf{Y}_{2} = \textbf{Z}_2 + \textbf{n}_2,   
\end{equation}
where $\textbf{n}_2 \sim \mathcal{CN} (0, {\sigma^{2}} )$.
At the receiver, both $\textbf{Y}_{1}$ and $\textbf{Y}_{2}$ are converted to real vectors.
The channel signal-to-noise ratio (SNR) between Alice and the legitimate user is the same as that between Alice and the eavesdropper.

Both Bob and Eve share the goal of reconstructing the source image $\textbf{X}$ as accurately as possible.
For the legitimate user, we first divide $\textbf{Y}_{1}$ into $\textbf{Y}_{\rm private}^1$ and $\textbf{Y}_{\rm common}^1$ following the previous steps, 
and then feed $\textbf{Y}_{\rm private}^1$ into a NN-based DP deprotection module to remove the fake DP noise, denoted by
\begin{equation}
\hat{\textbf{Y}}_{\rm private}^{1} = f_{\mathrm{deprotection}}\left( {\textbf{Y}_{\rm private}^1;\bm{\theta}_{1}^{\mathrm{deprotection}}} \right),
\end{equation}
where $f_{\mathrm{deprotection}}$ represents the NN-based DP deprotection module.
%
%
Subsequently, we combine $\hat{\textbf{Y}}_{\rm private}^{1}$ with $\textbf{Y}_{\rm common}^1$ to obtain the semantic information $\textbf{S}_{1}$, denoted by
\begin{equation}
    \textbf{S}_1 = \left[\bar{\textbf{Y}}_{\rm private}^{1}, \textbf{Y}_{\rm common}^1\right].
\end{equation}

Finally, the legitimate user inputs $\textbf{S}_1$ into a pretrained semantic StyleGAN generator $f_{\mathrm{gen}}$ to obtain the reconstructed image $\hat{\textbf{X}}_{1}$, denoted by
\begin{equation}
    \hat{\textbf{X}}_{1} = f_{\mathrm{gen}}\left( {\textbf{S}_{1}} \right).
\end{equation}
For the eavesdropper,
she directly inputs $\textbf{Y}_2$ into $f_{\mathrm{gen}}$ to get the reconstructed image $\hat{\textbf{X}}_2$, denoted by
\begin{equation}
    \hat{\textbf{X}}_{2} = f_{\mathrm{gen}}\left( {\textbf{Y}_2} \right).
\end{equation}

\subsection{Proposed privacy-preserving SemCom }
Our proposed privacy-preserving SemCom system can be divided into four parts, 
including the inversion method of the semantic StyleGAN, 
the NN-based DP protection / deprotection modules, 
the approximate privacy guarantee, 
and the calculation of the sensitivity $\Delta f$ in DP. 
%

\subsubsection{The Inversion Method of the Semantic StyleGAN}

The process of the inversion method is as follows. We start with an initial latent representation and input it into the semantic StyleGAN generator to obtain the generated image. 
Next, we calculate the mean square error (MSE) loss between the generated image and the source image, iterating the process to minimize this loss. 
Finally, after a certain number of iterations, we obtain the optimal latent representation.

\subsubsection{NN-based DP Protection / Deprotection Modules}

\begin{figure}[htbp]
\begin{center}
\centerline{\includegraphics[width=1\linewidth]{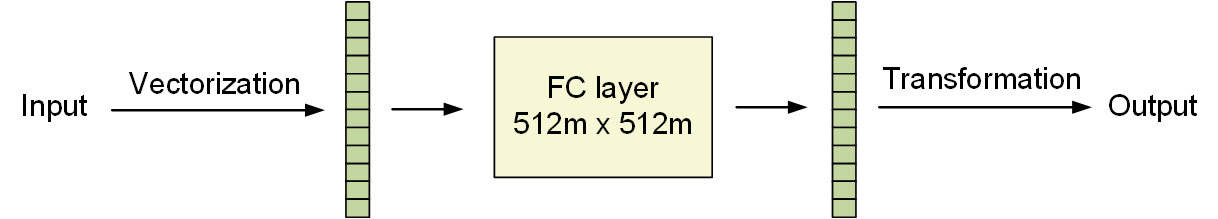}}
\caption{The structure of the NN-based DP protection module.}
\label{DP pro}
\end{center}
\vskip -0.3in
\end{figure}

The network structure of the NN-based DP protection module is shown in Fig.~\ref{DP pro}.
The DP protection module learns to add fake DP noise to $\textbf{Z}_{\rm private} \in \mathbb{R}^{m \times 512}$ that is similar to $\textbf{n}_{\rm dp}$. This module consists of a fully connected (FC) layer.
The number $ N_{\rm input} \times N_{\rm output}$ below the FC layer represents its configuration, 
where $N_{\rm input}$ and $N_{\rm output}$ are the numbers of input and output neurons, respectively.
Specifically, the DP protection module first vectorizes the input private latent codes to $\textbf{Z}_{\rm private}^{'} \in \mathbb{R}^{512m}$.
Then it inputs $\textbf{Z}_{\rm private}^{'}$ into a FC layer and transforms the output to $\hat{\textbf{Z}}_{\rm private}^{2}\in \mathbb{R}^{m \times 512}$.
The DP deprotection module learns to remove the fake DP noise, and its network structure is consistent with that of the DP protection module. 
%
%

\subsubsection{The Approximate Privacy Guarantee}
%

We know that the genuine DP noise added to $\textbf{Z}_{\rm private}$ is $\textbf{n}_{\rm dp} \sim Lap(0, \frac{\Delta f}{\epsilon})$. 
We then calculate the fake DP noise added to $\textbf{Z}_{\rm private}$ by the NN-based DP protection module, denoted as
\begin{equation}
\hat{\textbf{n}}_{\rm dp} = \hat{\textbf{Z}}_{\rm private}^{2}-\textbf{Z}_{\rm private}.
\end{equation}
To approximate the DP mechanism, we need to ensure that $\hat{\textbf{n}}_{\rm dp}$ approximately follows some Laplace distribution. 
Thus, we fit the probability distribution of $\hat{\textbf{n}}_{\rm dp}$ to the closest Laplace probability distribution.
Specifically, the fitted Laplace distribution $Lap(0, s)$ has a location parameter of 0 and a scale parameter of $s$.
We assume that the NN-based DP protection module provides an approximate $\epsilon'$-DP guarantee, then we have
\begin{equation}
s= \frac{\Delta f}{\epsilon'},\epsilon'= \frac{s}{\Delta f}.
\end{equation}
Since we know $s$ and $\Delta f$, the NN-based DP protection module provides an approximate $\frac{s}{\Delta f}$-DP guarantee. 
Furthermore, in general, we have $\epsilon'>\epsilon$ because the DP protection module also considers the reconstruction performance of the legitimate user as it learns to add fake DP noise, thus approximately providing weaker DP guarantee.

\subsubsection{The Calculation of the Sensitivity in DP}

With regard to DP for image protection \cite{xue2021dp}, 
the sensitivity refers to the maximum difference between the latent codes of any two different images in the training dataset. 
Specifically, we have two different images, $\textbf{I}_1$ and $\textbf{I}_2$, both from the training dataset. 
Let $f_{\rm inv}$ denote the inversion method of the semantic StyleGAN. 
Then, $\Delta f$ can be defined as
\begin{equation}
   \Delta f \stackrel{.}{=} \sup_{\textbf{I}_1, \textbf{I}_2 \in \mathcal{D}} \left\| f_{\rm inv}(\textbf{I}_1) - f_{\rm inv}(\textbf{I}_2) \right\|_2,
   \label{sensi}
\end{equation}
where $\mathcal{D}$ is the training dataset.
To facilitate the calculation of $\Delta f$ and to remove outliers in the latent codes, we clip the latent codes to retain 99\% of the features and calculate the sensitivity based on the clipped latent codes.
The clipping process is as follows. 
First, we transform each image in the training dataset into latent codes. 
Then, we calculate the 0.5\% quantile $a$ and the 99.5\% quantile $b$ of all the latent codes.
Finally, we adjust the elements in the latent codes that are out of the range of $\left[a,b\right]$ to either $a$ or $b$. 
After clipping, we calculate the sensitivity $\Delta f$ as follows:
\begin{equation}
\Delta f = \left\| b\mathbb{I}_n - a\mathbb{I}_n \right\|_2  = \sqrt{(b - a)^{2} \cdot n},
\end{equation}
where $\mathbb{I}_n$ is an all-ones vector of length $n$, and $n$ is the number of elements in the latent codes of an image.

\subsection{Training Strategy}

The performance gap between the legitimate user and the eavesdropper relies on the DP protection module and the privacy budget.
Since we use the pretrained semantic StyleGAN, we do not optimize the process of generating the latent codes $\textbf{Z}$ and the reconstructed image $\hat{\textbf{X}}$. 
Instead, we focus on how to reliably transmit the latent codes $\textbf{Z}$ to the receiver.
The loss function can be written as
\begin{equation}
    \mathcal{L} = MSE\left( {\textbf{Z},\textbf{S}_{1}} \right) + \lambda \cdot MSE\left( {\hat{\textbf{Z}}_{\rm private}^{1},\hat{\textbf{Z}}_{\rm private}^{2}} \right),
\end{equation}
where the first term measures the image reconstruction performance of the legitimate user, and the second term measures the performance of the DP protection module in adding DP noise. 
Here, $\lambda$ is the trade-off hyperparameter.

\section{Simulation Results}

\subsection{Experimental Settings}


In this subsection, we demonstrate our experimental settings.

\textbf{Pretrained Model and Dataset}:
The pretrained model we use is a semantic StyleGAN model trained on the CelebAMask-HQ dataset \cite{lee2020maskgan}, where each image in the dataset is resized to $512 \times 512$.
We use the CelebAMask-HQ dataset for training and testing, and all these images are resized to $512 \times 512$ too.

\textbf{Private Latent Codes}: 
The dimension of the latent codes for each image is $28 \times 512$, where 28 is the number of latent codes.
The first two latent codes are shared latent codes, and the rest are shape and texture codes (local latent codes), all of which have a length of 512.
Shared latent codes are always private latent codes, in addition we select some latent codes from local latent codes as private latent codes.

\textbf{Privacy Budget and Channel SNR}:
The value of the sensitivity $\Delta f$ is 351.88, and the privacy budget $\epsilon$ falls within the set $\epsilon \in \{1,3,5,8,10,15,30,100,300,800\}$.
The channel SNR is 20dB for both Bob and Eve.

\textbf{Training Settings}:
The batch size is 512, and we use the cosine annealing warm restarts optimizer to train our proposed SemCom system.
We set $\lambda = 1\times 10^{-3}$.
The shared latent codes and the 4th to 8th latent codes are private latent codes.
The initial learning rate is $3 \times 10^{-4}$, and the training period is 100 epochs.

\textbf{Performance Metrics}:
We utilize the learned perceptual image patch similarity (LPIPS) \cite{zhang2018unreasonable} as a performance metric.
Specifically, we adopt the Alexnet-based LPIPS \cite{krizhevsky2012imagenet}.
In addition, we use the face privacy protection success rate (FPPSR) as a performance metric. 
Specifically, we use a face recognition system called ArcFace \cite{deng2019arcface} to determine whether the input image pairs are the same person. 
The FPPSR refers to the percentage of received faces that are determined to be different from the original faces.



\textbf{The Benchmark}:
%
In the benchmark, Alice transmits the latent codes $\textbf{Z}$ directly to the legitimate user without using the DP protection module. 
The image reconstruction performance of the legitimate user and the eavesdropper is the same since they have the same channel SNR. 
We name this benchmark as \textit{Direct Transmission without Protection}. 

%






\subsection{Experimental Results}

\begin{figure}[!t]
\begin{center}
\centerline{\includegraphics[width=0.8\linewidth]{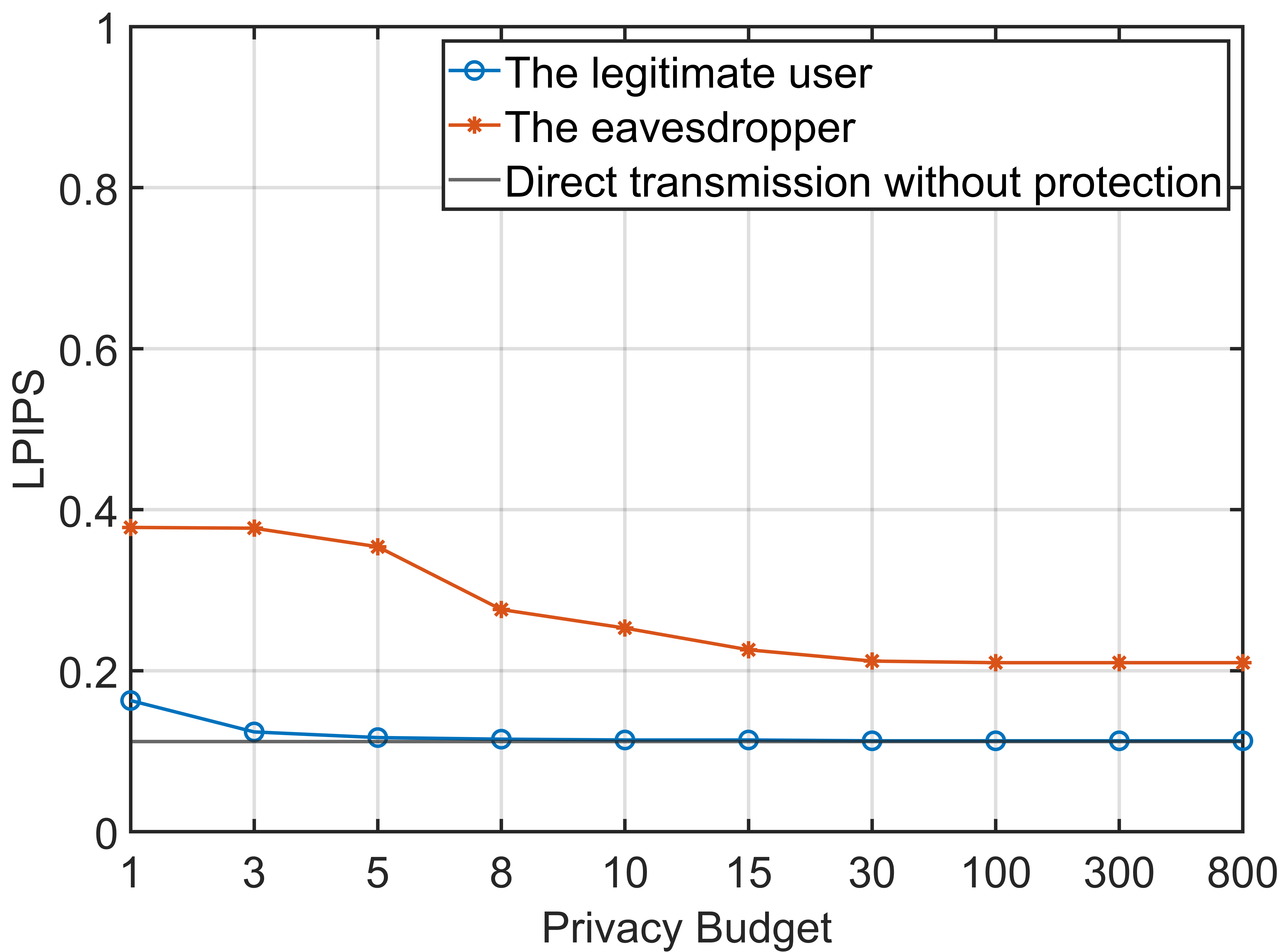}}
\caption{The LPIPS performance of the proposed system under different privacy budgets $\epsilon$ when SNR is 20dB.}
\label{fig.LPIPS1}
\end{center}
\vskip -0.3in
\end{figure}

In this subsection, we evaluate the security and performance of the proposed system under different privacy budgets with SNR = 20dB. 
The lower the LPIPS / FPPSR of the legitimate user, the better the image reconstruction performance of the system, and the higher the LPIPS / FPPSR of the eavesdropper, the better the security of the system. 
Specifically, we first examine the LPIPS performance of the legitimate user and the eavesdropper, as shown in Fig.~\ref{fig.LPIPS1}.
From Fig.~\ref{fig.LPIPS1}, it can be observed that the LPIPS performance of both the legitimate user and the eavesdropper increases with the privacy budget. 
The LPIPS performance of the legitimate user reaches the upper bound, i.e., the LPIPS performance of \textit{Direct Transmission without Protection}, when $\epsilon \geq 5$. 
This indicates that the DP deprotection module can effectively remove the privacy-preserving noise.
The LPIPS performance of the eavesdropper, on the other hand, is always lower than that of the legitimate user, with a gap of about 0.1-0.2 in the LPIPS values. This indicates that the proposed DP protection module can effectively improve the security of the system.
%


\begin{figure}[!t]
\begin{center}
\centerline{\includegraphics[width=0.8\linewidth]{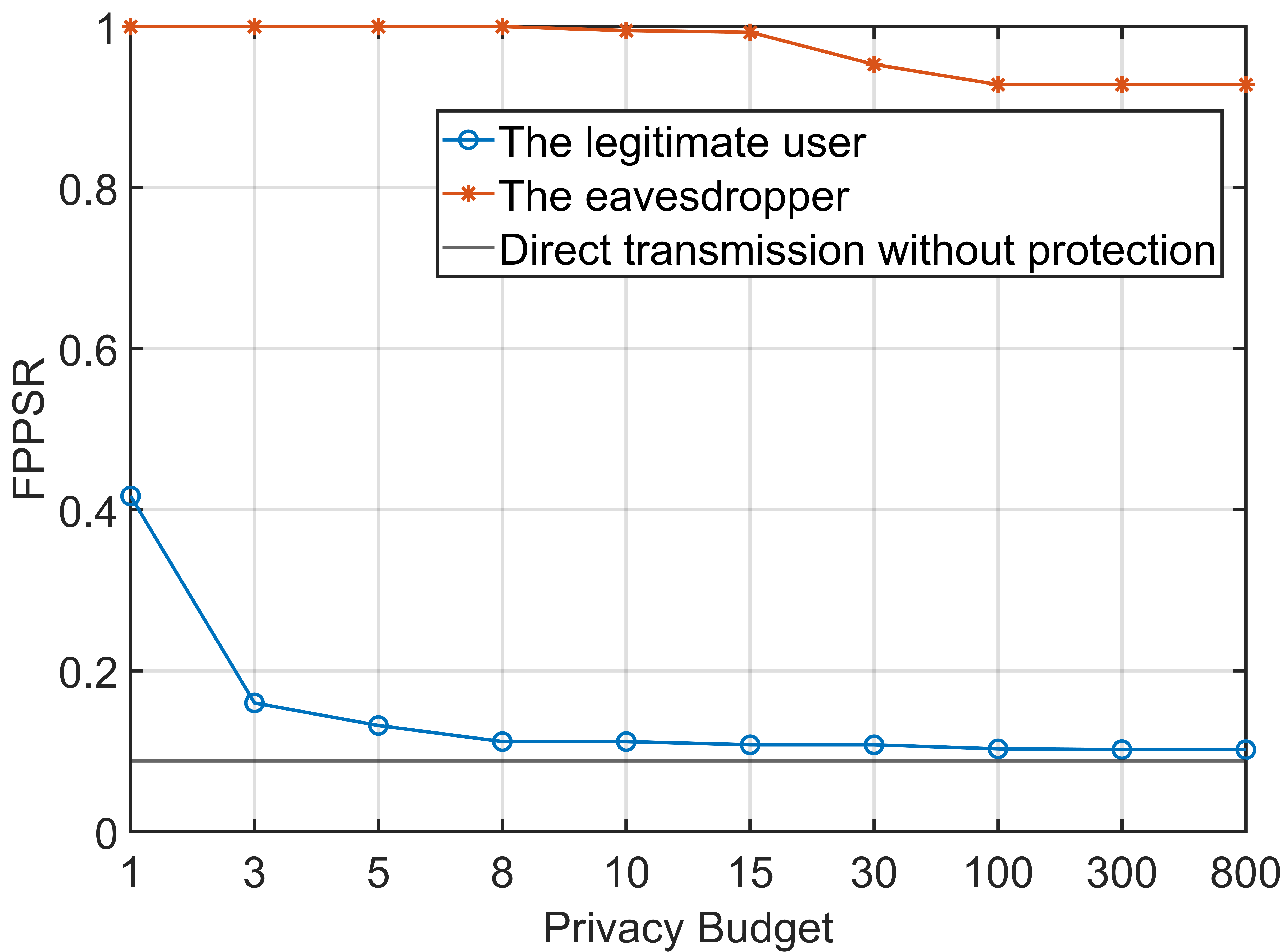}}
\caption{The FPPSR performance of the proposed system under different privacy budgets $\epsilon$ when SNR is 20dB.}
\label{fig.FPPSR1}
\end{center}
\vskip -0.3in
\end{figure}

\begin{figure*}[htbp]
\centering
\subfigure[Source image]{
\begin{minipage}[t]{0.16\linewidth}
\centering
\includegraphics[width=1.1in]{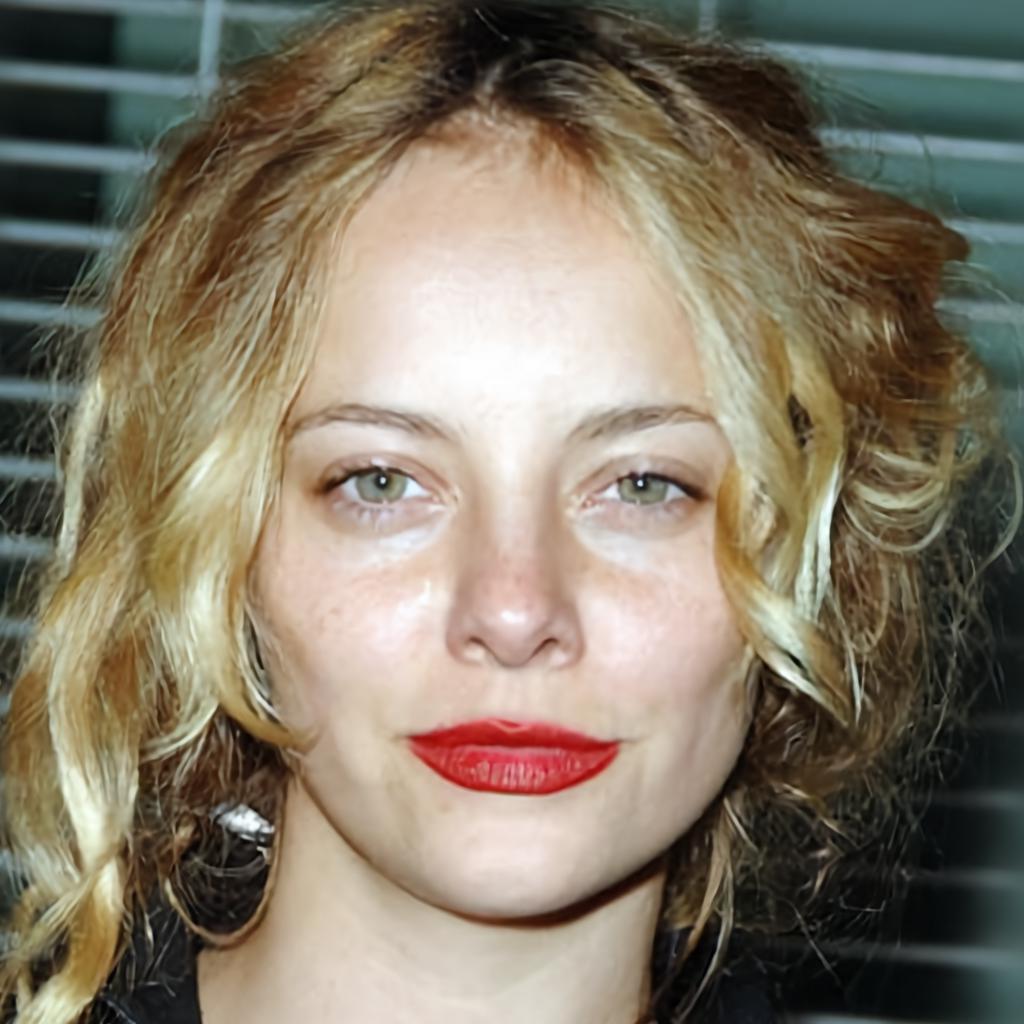}
\end{minipage}%
}%
\subfigure[$\epsilon = 1$]{
\begin{minipage}[t]{0.16\linewidth}
\centering
\includegraphics[width=1.1in]{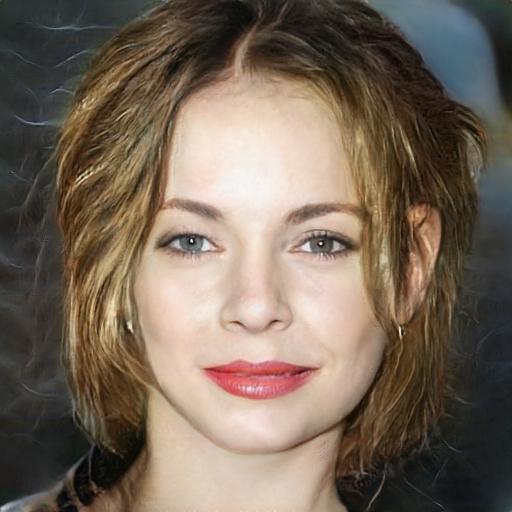}
\end{minipage}%
}%
\subfigure[$\epsilon = 10$]{
\begin{minipage}[t]{0.16\linewidth}
\centering
\includegraphics[width=1.1in]{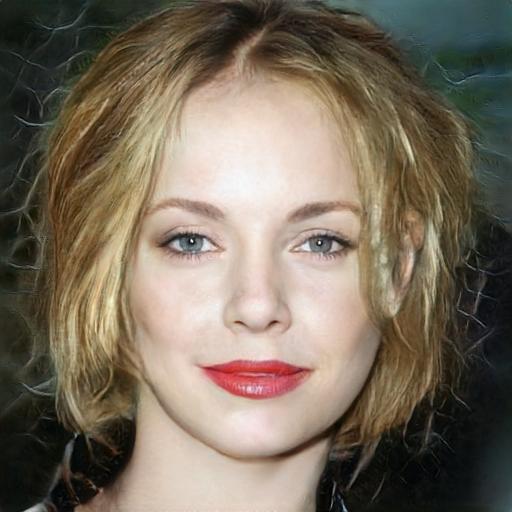}
\end{minipage}
}%
\subfigure[$\epsilon = 15$]{
\begin{minipage}[t]{0.16\linewidth}
\centering
\includegraphics[width=1.1in]{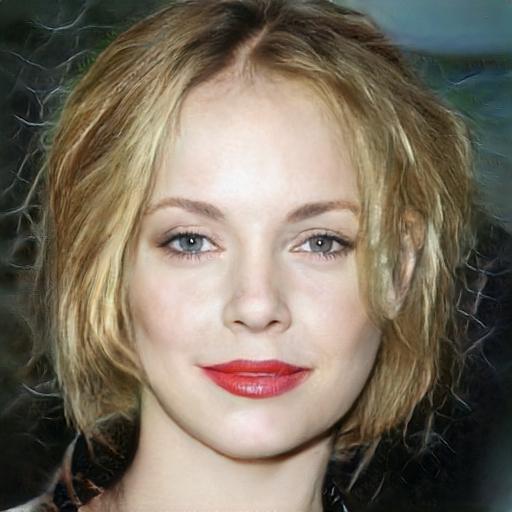}
\end{minipage}
}%
\subfigure[$\epsilon = 300$]{
\begin{minipage}[t]{0.16\linewidth}
\centering
\includegraphics[width=1.1in]{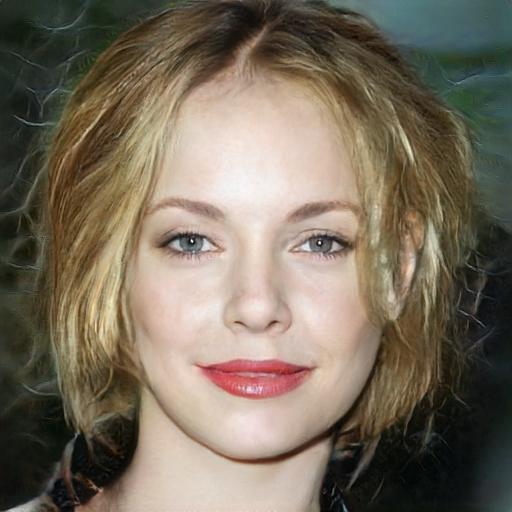}
\end{minipage}
}%
\subfigure[$\epsilon = 800$]{
\begin{minipage}[t]{0.16\linewidth}
\centering
\includegraphics[width=1.1in]{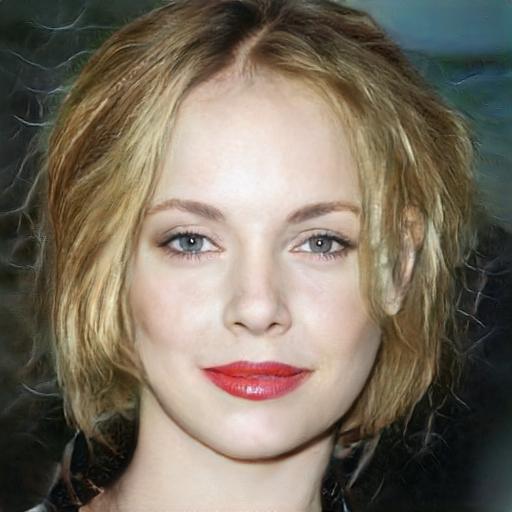}
\end{minipage}
}%
\centering
\caption{Visual analysis of the reconstructed images by the legitimate user under different privacy budgets $\epsilon$ when SNR is 20dB.}
\label{fig.visual1}
\vskip -0.2in
\end{figure*}

\begin{figure*}[htbp]
\centering
\subfigure[Source image]{
\begin{minipage}[t]{0.16\linewidth}
\centering
\includegraphics[width=1.1in]{with_dp_only_2/original_image.jpg}
\end{minipage}%
}%
\subfigure[$\epsilon = 1$]{
\begin{minipage}[t]{0.16\linewidth}
\centering
\includegraphics[width=1.1in]{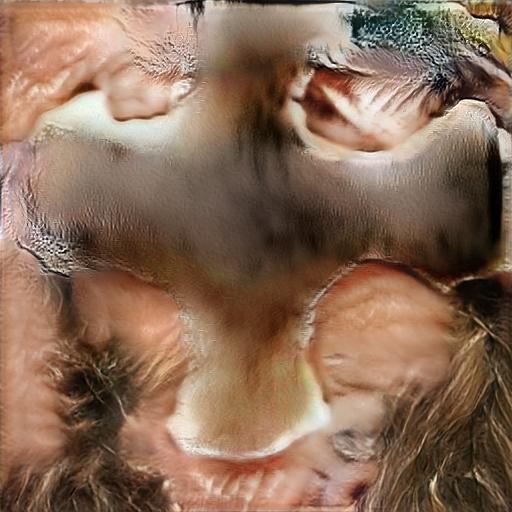}
\end{minipage}%
}%
\subfigure[$\epsilon = 10$]{
\begin{minipage}[t]{0.16\linewidth}
\centering
\includegraphics[width=1.1in]{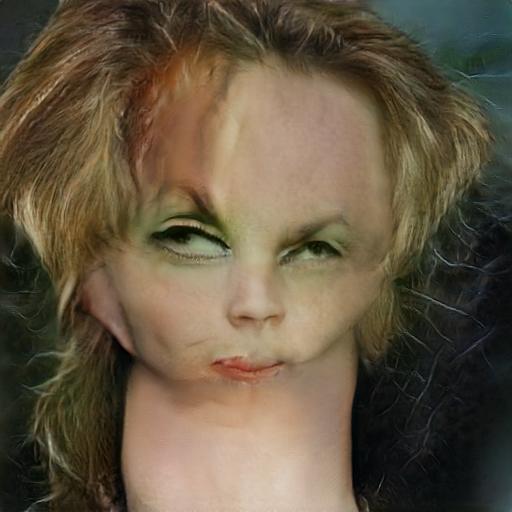}
\end{minipage}
}%
\subfigure[$\epsilon = 15$]{
\begin{minipage}[t]{0.16\linewidth}
\centering
\includegraphics[width=1.1in]{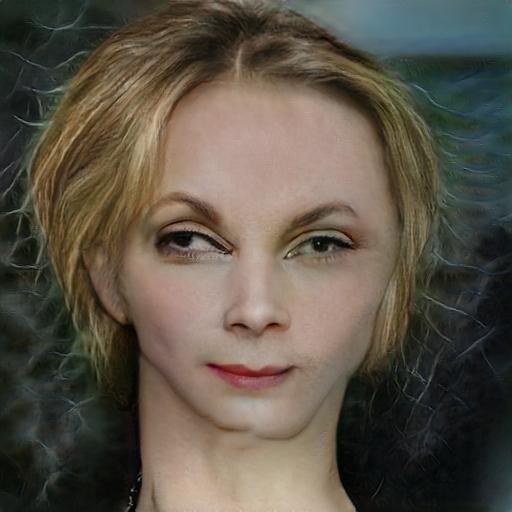}
\end{minipage}
}%
\subfigure[$\epsilon = 300$]{
\begin{minipage}[t]{0.16\linewidth}
\centering
\includegraphics[width=1.1in]{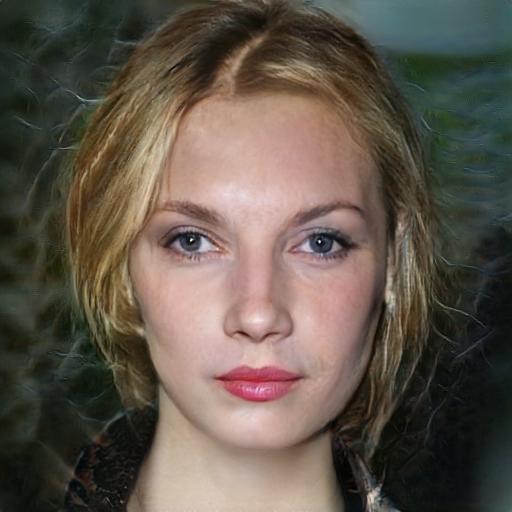}
\end{minipage}
}%
\subfigure[$\epsilon = 800$]{
\begin{minipage}[t]{0.16\linewidth}
\centering
\includegraphics[width=1.1in]{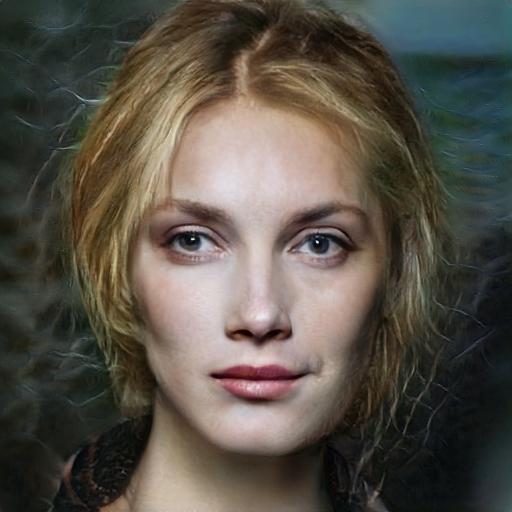}
\end{minipage}
}%
\centering
\caption{Visual analysis of the reconstructed images by the eavesdropper under different privacy budgets $\epsilon$ when SNR is 20dB.}
\label{fig.visual2}
\vskip -0.2in
\end{figure*}

Secondly, we examine the FPPSR performance of the legitimate user and the eavesdropper, as shown in Fig.~\ref{fig.FPPSR1}. 
From Fig.~\ref{fig.FPPSR1}, we can see that the FPPSR performance of both the legitimate user and the eavesdropper increases with the privacy budget, but the increase in the FPPSR performance of the eavesdropper is very limited. 
The FPPSR performance of the legitimate user is very close to the upper bound, indicating that the system has excellent performance. 
The FPPSR performance of the eavesdropper is always poor, with the FPPSR value is close to 0.9-1. 
This indicates that the system has a high level of security and can effectively prevent eavesdroppers from obtaining valid information.

Finally, we visualize the reconstructed images of the legitimate user and the eavesdropper under different privacy budgets with SNR = 20dB to evaluate the effectiveness of the proposed method in terms of image protection and deprotection. 
The reconstructed images of the legitimate user and the eavesdropper are shown in Fig.~\ref{fig.visual1} and Fig.~\ref{fig.visual2}, respectively.
From Fig.~\ref{fig.visual1}, we can find that when $\epsilon = 1$, 
the gap between the reconstructed image of the legitimate user and the source image is relatively large, 
with differences in the eye and nose areas. 
As the privacy budget $\epsilon $ increases to 10 and above, the reconstructed images of the legitimate user become highly similar to the source image. 
From Fig.~\ref{fig.visual2}, we can find that when $\epsilon = 1$, the reconstructed image of the eavesdropper is completely chaotic, making it almost impossible for the eavesdropper to obtain any valid information. 
This is due to the strong privacy protection provided for the latent codes of the image, leading the eavesdropper to synthesize confusing semantics when generating the source image. 
As $\epsilon$ increases, the eavesdropper is able to generate clear face images (e.g., $\epsilon = 15,300,800$), but the gap between the reconstructed image and the source image is so large that it is visually impossible to tell that the reconstructed image and the source image are the same person. 
This indicates that our approach can obfuscate images or generate fake images by adjusting the privacy budget to achieve different security levels.

%

\section{Conclusion}
In this paper, we introduced a DP based secure SemCom system over a wiretap channel. 
Our proposed method offers several significant advantages. Firstly, compared to existing DP schemes for image protection, the proposed method enables the legitimate user to effectively remove privacy protection. 
Secondly, compared to other secure SemCom systems, the proposed method provides approximate privacy guarantee and does not require that the eavesdropper's channel conditions are worse or that the legitimate network can be optimized against the eavesdropper's network. 
These advantages make our proposed method a practical and secure SemCom solution that can easily satisfy scenarios with different security requirements.



\vspace{12pt}
\end{CJK}
\end{document}